\documentclass[reprint, pdftex, amsmath,amssymb,
 aps,pra,showpacs,
]{revtex4-1}

\usepackage{graphicx}
\usepackage{dcolumn}
\usepackage{bm}
\usepackage{color}
\usepackage[normalem]{ulem}

\begin{document}

\preprint{APS/123-QED}

\title{Non-linear Spectroscopy of Sr Atoms in an Optical Cavity for Laser Stabilization}

\author{B. T. R. Christensen$^1$}
 \email{bjarkesan@nbi.ku.dk}
\author{M. R. Henriksen$^1$}%
\author{S. A. Sch\"{a}ffer$^1$}%
\author{P. G. Westergaard$^{2}$}%
\author{J. Ye$^3$}%
\author{M. J. Holland$^3$}%
\author{J. W. Thomsen$^1$}%

\affiliation{%
$^1$Niels Bohr Institute, University of Copenhagen, Blegdamsvej 17, 2100 Copenhagen, Denmark\\
$^2$Danish Fundamental Metrology,  Matematiktorvet 307, 1. sal, 2800 Kgs. Lyngby, Denmark\\
$^3$JILA, National Institute of Standards and Technology and University of Colorado, Boulder, CO 80309-0440, USA
}%

\begin{abstract}
We study the non-linear interaction of a cold sample of strontium-88 atoms coupled to a single mode of a low finesse optical cavity in the so-called bad cavity limit and investigate the implications for applications to laser stabilization. The atoms are probed on the weak inter-combination line $|5s^{2} \, ^1 \textrm{S}_0 \rangle \,-\, | 5s5p \, ^3 \textrm{P}_1 \rangle$ at 689 nm in a strongly saturated regime. Our measured observables include the atomic induced phase shift and absorption of the light field transmitted through the cavity represented by the complex cavity transmission coefficient.  We demonstrate high signal-to-noise-ratio measurements of both quadratures -- the cavity transmitted phase and absorption -- by employing FM spectroscopy (NICE-OHMS). We also show that when FM spectroscopy is employed in connection with a cavity locked to the probe light, observables are substantially modified compared to the free space situation where no cavity is present. Furthermore, the non-linear dynamics of the phase dispersion slope is experimentally investigated and the optimal conditions for laser stabilization are established. Our experimental results are compared to state-of-the-art cavity QED theoretical calculations.

\begin{description}
\item[PACS numbers]32.70.Jz,32.80.Wr,37.30.+i,42.50.Ct,42.62.Fi
\end{description}
\end{abstract}

\maketitle

\section{Introduction}
Lasers with exceedingly pure spectral features are a key element in the interrogation of ultra narrow atomic transitions applied in for example optical atomic clocks \cite{BenBloom,Nicholson,Katori2015,Hinkley,Targat}, condensed matter simulations with cold atoms \cite{MartinScience}, and relativistic geodesy \cite{Bondarescu, Guena}. Such pure spectral instruments with high phase stability and long coherence times are essential to investigations of physics beyond the standard model, e.g., the drift of fundamental physical constants \cite{Rosenband,Godun} or the detection of gravitational waves \cite{Graham} as predicted by the general theory of relativity. The stability of the current state-of-the-art lasers are limited by Brownian motions of the reference cavity mirror substrates and coatings \cite{Kessler, Kessler2, Cole}, and the emergence of new technologies \cite{ThompsonSr} seems necessary if further improvements are to be made commonly available.

Recent studies have proposed a new scheme for laser stabilization to a mK thermal sample of atoms placed in a low finesse optical cavity \cite{PGW,DT,Meiser1,MikeMartin}. By placing the atoms inside the optical cavity atom-light interactions are enhanced by order of the finesse of the cavity and non-linear effects are brought into play, which enhances the spectral sensitivity of the detected photo current used for stabilization. Interestingly, the first proposals considered atoms trapped in an optical lattice or systems with a sample temperature equivalent to zero degrees Kelvin \cite{MikeMartin}. Further studies have extended these results to also include the finite temperature of the sample \cite{PGW,DT}. At finite temperatures the bi-stability regime, present for zero-Kelvin systems, turns out to be effectively removed. Since the optimal locking point for these systems is located at an input power range within the (zero Kelvin) bi-stability regime this sparks renewed interest for stabilization to samples of atoms with finite temperatures.

In this work we investigate the non-linear coupling of laser cooled strontium-88 atoms to a single mode of an optical cavity. During experiments the atom-modified cavity mode is forced to be on resonance with the carrier of the probe laser and thus a standing wave will be present at all times in the cavity.
As a result significant changes appear in the observables of the system due to reaction of the atoms on the cavity servo system.

This has major consequences: the phase and absorption information is altered compared to conventional free space FM spectroscopy, but may be recovered under special conditions of laser detuning and demodulation frequency. In the following we also explore the non-linear cavity-atom dynamics and identify optimal parameters relevant for laser stabilization to our cavity system.

\section{Experimental Setup}
\begin{figure}[ht]
\includegraphics[width=0.48\textwidth]{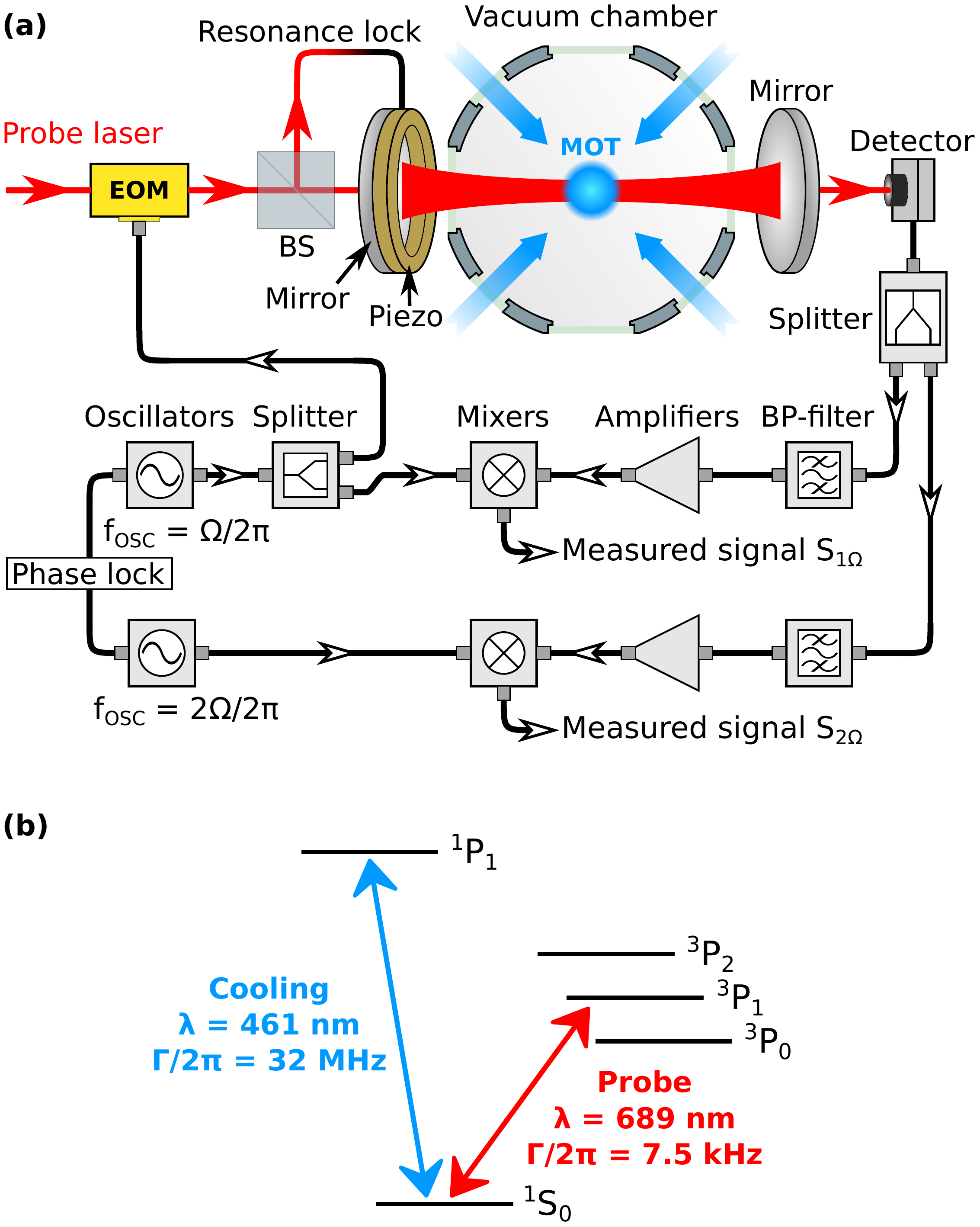}
\caption{{\bf (a)} Experimental setup. A sample of cold $^{88}$Sr atoms trapped in a standard magneto-optical trap (MOT) is located inside a low finesse optical cavity ($F=85$). The cavity is locked to resonance with the light using the H\"{a}nsch-Couillaud \cite{Hansch} scheme. We perform cavity-enhanced FM spectroscopy (NICE-OHMS) where the light is modulated at $\Omega/2\pi = \textrm{FSR}$ (Free spectral range of the cavity). The detected signal is split into two arms where the $\Omega$ and $2 \Omega$ components are selected by band-pass filters and separately demodulated using RF-mixers. One part of the signal is demodulated at $\Omega$ giving a phase signal and the other part is demodulated at $2 \Omega$ giving an attenuation signal. Both the phase shift and absorption lineshape of the transmitted probe light are recorded simultaneously with this scheme. {\bf (b)} Energy levels of the $^{88}$Sr atom important for this work. The singlet transition $|5s^{2} \, ^1 \textrm{S}_0 \rangle \,-\, | 5s5p \, ^1 \textrm{P}_1 \rangle$ at $461$~nm is used to trap and cool the atoms. The narrow intercombination line transition $|5s^{2} \, ^1 \textrm{S}_0 \rangle \,-\, | 5s5p \, ^3 \textrm{P}_1 \rangle$ at $689$~nm is used to probe the atoms.
\label{setup}}
\end{figure}

Experimentally, our cavity system consists of a sample of laser cooled strontium-88 atoms with a temperature of 5 mK coupled to an optical cavity with finesse $F=85$, see Fig.~\ref{setup}(a). We operate a standard Magneto-Optical Trap (MOT) and trap about $5\cdot 10^8$ strontium atoms using the strong $^{1}$S$_{0}\rightarrow^{1}$P$_{1}$ transition at $461\,\textrm{nm}$, see Fig.~\ref{setup}(b). The system is probed on the $^{1}$S$_{0}\rightarrow^{3}$P$_{1}$ intercombination transition at $689\,\textrm{nm}$ while the cavity is forced to be on resonance with the probe laser at all times. The cavity mirrors have a radius of curvature of $9$~m yielding a cavity waist diameter of about $1$~mm which is comparable to the size of the MOT. At our current MOT temperature we estimate the transit time broadening to be $1-2$~kHz which is significantly smaller than the natural atomic linewidth of $\Gamma/2\pi = 7.5$~kHz. The $30$~cm cavity, with a free spectral range (FSR) of $500$~MHz, is placed outside the vacuum chamber thereby limiting the obtainable finesse. We measure a cavity linewidth of $\kappa/2\pi = 5.8$~MHz placing our system in the bad cavity regime $\kappa>>\Gamma$.
We probe the atoms for $100\,\mu\textrm{s}$ and the $461\,\textrm{nm}$ light is turned off during this probing period. At the probing time scale our pre-stabilized laser has an estimated linewidth of $800$~Hz. 

We employ cavity-enhanced FM spectroscopy (NICE-OHMS)\cite{NICEOHMS,NICEOHMS2}, see Fig.~\ref{setup}, where the probe carrier is modulated at the FSR to create sidebands, in our case separated from the carrier by multiples of $500$~MHz. The cavity transmitted signal is detected by a high bandwidth photo-detector recording the beat note between the carrier and the sidebands. This signal contains information about the cavity transmitted field amplitude and non-linear phase shifts due to the atoms present in the cavity. These contributions are recovered with a high signal-to-noise-ratio by demodulating at FSR and multiple FSRs. 

Our cavity system supports a single atom cooperativity $C_0 = 4g^2/\Gamma\kappa$ of $C_0 = 3\cdot 10^{-5}$, where $g$ is the single-atom-cavity coupling constant. For our cavity setup we find $g/2\pi=$ 590 Hz. The $C_0$ parameter is a measure of the rate at which an atom emits a photon into the cavity mode compared to all other dissipation rates in the system. For the collective cooperativity we find \mbox{$C = N_{\textrm{\tiny{cav}}}\cdot C_0= 630$}, where $N_{\textrm{\tiny{cav}}}$ is the total number of atoms in the cavity mode volume. With a MOT atom number of $5\cdot 10^8$ we find $ N_{\textrm{\tiny{cav}}}= 2.5\cdot 10^7$.

\section{Theory of measurement}

In this section we establish a connection between our measured cavity transmitted quantities and the complex transmission coefficient of the input field $\chi$.
Our input probe field can be written in terms of a carrier and multiple orders of sidebands induced by phase-modulation. By neglecting residual amplitude modulation, the total input probe field, $E_{\textrm{\tiny{in}}}$, with sidebands can be given as 
\begin{eqnarray}
E_{\textrm{\tiny{in}}}&=& E_0 \sum_{p=-\infty}^\infty J_p(x) e^{i(\omega_c + p\Omega t)} ,\label{components}
\end{eqnarray}

where $J_p(x)$ is the $p$'th order regular Bessel function describing the amount of power transferred from the carrier ($p=0$) to the sidebands ($|p|>0$), $x$ is the EOM modulation index, $\omega_c $ is the carrier frequency which is close to the atomic resonance and $\Omega$ is the EOM modulation frequency equal to the free spectral range of the cavity. Each transmitted component of the cavity field is modified by the cavity according to the well known expression \cite{Riehle},
\begin{eqnarray}
E_{\textrm{\tiny{out}}}&=& \frac{Te^{i\varphi}}{1-R e^{i2\varphi}}E_{\textrm{\tiny{in}}}=\chi E_{\textrm{\tiny{in}}},
\end{eqnarray}
where $\chi$ is the complex transmission coefficient describing the phase change and the absorption of the field, $T$ $(R)$ is the power transmission (reflection) coefficient of the cavity mirrors, and $\varphi$ is the single round-trip complex phase picked up by the intra-cavity field.

The carrier frequency is used to probe the atomic transition and is close to the atomic resonance while the sidebands at $\Omega/2\pi=500\,\textrm{MHz}$ are far off resonance compared to the natural linewidth, $\Gamma/2\pi= 7.5$~kHz. Each of the field's phase components -- carrier ($\varphi_0$) and sidebands ($\varphi_{\pm p}$ for $p=1,2$) -- experiences identical phase shifts due to the cavity assembly $\phi_{\textrm{\tiny cavity}}$, but only the carrier interacts with the atoms,
\begin{equation}
\varphi_0 = \phi_{\textrm{\tiny{cavity}}}+ \beta_{D}+i\beta_{A},\quad\varphi_{\pm p}=\phi_{\textrm{\tiny{cavity}}} \pm p\pi,
\end{equation}
where $\beta_{D}$ and $\beta_{A}$ denotes the single pass dispersion and absorption due to the atoms present in the cavity, and the sideband frequency detuning of $\pm p\Omega$ introduces additional phase factors of $\pm p \pi$, if the cavity is on resonance with the carrier and the sidebands.
During experiments the cavity is forced to be on resonance with the probe laser, a requirement that generally imposes restrictions on measured quantities in cavity experiments. Enforcing the resonance condition by locking the cavity to the probe carrier implies
\begin{equation}
\Re (\varphi_0) = \theta_0 = \phi_{\textrm{\tiny{cavity}}}+ \beta_{D} = m\pi,\label{resonancecondition}
\end{equation}
where $\theta_0$ is the real part of $\varphi_0$ and $m$ is an integer. This imposes restrictions on the sideband phases which become
\begin{equation}
\varphi_{\pm p}=\phi_{\textrm{\tiny{cavity}}}-\theta_0 \pm p \pi= -\beta_{D} \pm p\pi.
\end{equation}
For the carrier the complex transmission coefficient $\chi_0$ becomes purely real
\begin{equation}
 \chi_0=\frac{Te^{-\beta_{A}}}{1-R e^{-2\beta_{A}}},\label{Chi0}
\end{equation}
while the sideband complex transmission coefficients $\chi_{\pm p}$ become
\begin{equation}
\chi_{\pm p} =\frac{Te^{i(\pm p \pi - \beta_{D})}}{1-R e^{2i(\pm p \pi -\beta_{D})}},\label{chip}
\end{equation}
and the atomic phase shift $\beta_{D}$ is transferred onto the sideband transmission coefficients.
As the cavity is kept on resonance with the carrier at all times we obtain the final output field to second order, $E_{\textrm{\tiny{out}}}$:


\begin{eqnarray}
E_{\textrm{\tiny{out}}}&=& E_0\Bigl(J_0(x) \chi_0 e^{i\omega_c t}\phantom{e^{i(\omega_c - 2\Omega)t}} \nonumber\\
& &+J_1(x) \chi_1 e^{i(\omega_c + \Omega)t}- J_1(x)\chi_{-1} e^{i(\omega_c - \Omega)t}\nonumber\\
& & +J_2(x) \chi_2 e^{i(\omega_c + 2\Omega)t}+ J_2(x)\chi_{-2} e^{i(\omega_c - 2\Omega)t}\Bigr).
\end{eqnarray}

Detecting this signal on a fast photodiode and demodulating at $\Omega/2\pi = \textrm{FSR}$ gives a photo current signal $S_{\scriptscriptstyle{1\Omega}}$ proportional to
\begin{eqnarray}
S_{1\Omega}&\propto&J_{0}(x)J_{1}(x)\chi_{0}\cdot \Im(\chi_{1}),\label{1omega}
\end{eqnarray}
where $\Im(\chi_{1})$ is the imaginary part of the $\chi_{1}$ coefficient.

In the limit $R\rightarrow 0$, corresponding to the free space case where no
cavity is used, we recover from Eq.~(\ref{1omega}) the well know expression from FM spectroscopy $S_{1\Omega}\propto e^{-\beta_{A}}\cdot\sin(\beta_{D})$. On the other hand in the limit where the complex phase shift due to the atoms is very small, i.e., $\beta_{D} \rightarrow 0$ and a cavity is present we find $S_{1\Omega}\propto \beta_{D}$, which is also proportional to the total atomic phase shift of the transmitted sideband field. In our case this approximation is satisfied very close to resonance.


Information related to the absorption of the transmitted light may be obtained by demodulating the transmitted field at twice the FSR, i.e., $2\Omega$, where one obtains (here with sidebands up to second order),
\begin{eqnarray}
S_{2\Omega}&\propto& \left( 2J_{0}(x)J_{2}(x)\chi_{0}\cdot \Re(\chi_{2})-J^{2}_{1}(x)|\chi_{1}|^{2} \right),\label{2omega}
\end{eqnarray}
where $\Re(\chi_{2})$ is the real part of the $\chi_{2}$ coefficient.
This signal is related to the absorption line shape of the transmitted field (see Appendix~\ref{appendix}). The complete information about the complex transmission coefficients, $\chi_j$, can now be found by combining the quantities obtained from demodulating at FSR and 2FSR via Eq.~(\ref{1omega}) and~(\ref{2omega}). 

In conclusion, the demodulating of the cavity transmitted field by $\Omega/2\pi = \textrm{FSR}$ (Eq.~\ref{1omega}) provides a signal proportional to the atomically induced phase shift for small detunings. The high signal-to-noise-ratio of this signal makes it promising for applications in laser stabilization. The absorption line shape on the other hand can be obtained by demodulating the cavity transmitted field by 2FSR (Eq.~\ref{2omega}). This technique can also be applied to molecular systems at ambient temperatures coupled to optical cavities and it is not only limited to laser cooled neutral atoms.

\section{Results and discussion}
\begin{figure}[h!t]
\includegraphics[width=1.0\columnwidth]{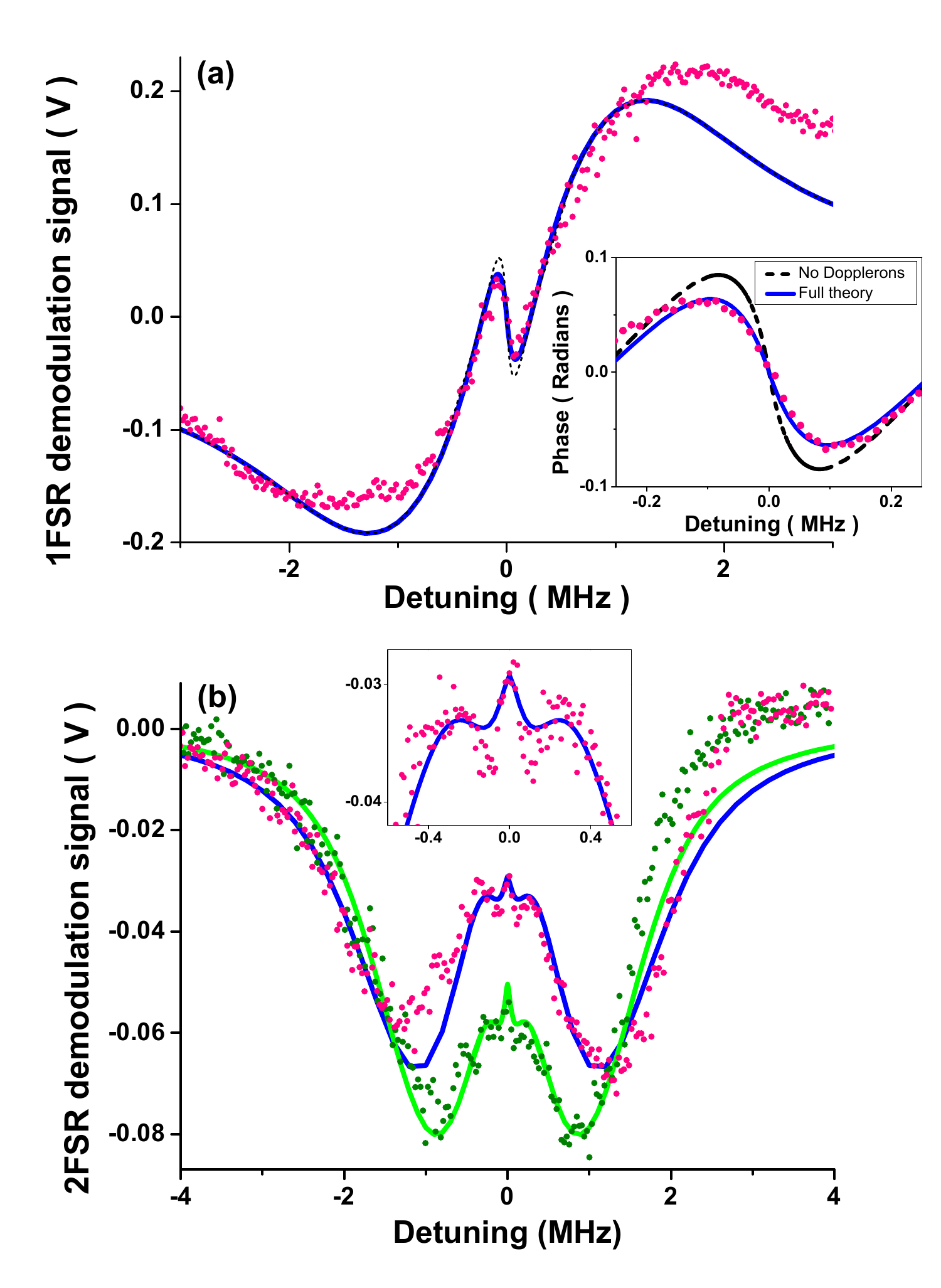}
\caption{The solid and dashed lines are predictions based on the theoretical model presented in \cite{PGW,DT}. The dots are data measured by cavity-enhanced FM spectroscopy. All data are averaged values, where approximately five times as many measurements have been performed for data in (b) due to the weaker signal compared to data in (a). 
The slight asymmetry of the experimental data in both plots is expected to be due to a drift of the relative phase between the light wavefront and the demodulation phase during scan. \textbf{(a)} Frequency scan of the cavity transmitted phase shift close to resonance for an carrier input power of $240\,\textrm{nW}$, total number of atoms of $N=5.0\cdot 10^{8}$ and atom temperature of $T=5.0\,\textrm{mK}$. The data points are averages of 9 measurements. 
Theoretical values from Eq.~(\ref{1omega}) are scaled to data. \textit{Inset}: High resolution scan of the central region. The vertical axis in this central region corresponds to the total atomic phase shift of the transmitted sideband and data is scaled to theory. 
\textbf{(b)} Frequency scans of the absorption lineshape of the cavity transmitted field measured by the NICE-OHMS technique obtained with a demodulation $\Omega/2\pi =$ 2FSR. The total number of atoms is $N=4.5\cdot 10^{8}$, atom temperature of $T=5.0\,\textrm{mK}$ and total input power of $310\,\textrm{nW}$ (blue line) and $155\,\textrm{nW}$ (green line). 
Theoretical values from Eq.~(\ref{2omega}) are scaled to data. \textit{Inset}: High resolution scans around the central region. The units on the inset axes are the same as in (b).
\label{Broadscans}}
\end{figure}
In our system the energy scale associated with the mK Doppler temperature is several orders of magnitude larger than the energy scale associated with the narrow linewidth of the optical transition. In this limit the non-zero velocities of the atoms bring additional multi-photon resonance phenomena into play, so-called doppleron resonances, which modify the complex transmission coefficient of the cavity field close to the atomic resonance \cite{PGW}.

Doppleron resonances are multiphoton processes where $l+1$ photons from one direction are  resonantly absorbed by an atom with a velocity $v$, and $l$ photons are emitted in the reverse direction leaving the atom in the excited state \cite{Tallet,Stenholm}. Energy conservation demands the doppleron resonances to be located approximately at \cite{Tallet}
\begin{eqnarray}
kv = \frac{\pm(\Delta^2 + 2g^2)^{1/2}}{2l+1}
\end{eqnarray}
where $k$ is the wave vector of the laser light, $v$ the resonant atom velocity, $\Delta$ the detuning and $l$ the number of emitted photons, i.e., the order of the doppleron resonance and $g$ is the single-atom-cavity coupling constant.

NICE-OHMS signals obtained by demodulation at $\Omega/2\pi =$ FSR, see  Eq.~(\ref{1omega}), are shown in Fig.~\ref{Broadscans}(a). Those values represent the total phase dispersion induced by the atom-cavity system. 
A sharp dispersion signal with a steep slope can be seen around resonance. The magnitude of this slope, as well as the signal-to-noise-ratio, ultimately determines the obtainable frequency stability of the system when used as a frequency discriminator for laser stabilization.
The inset in Fig.~\ref{Broadscans}(a) shows a zoom of the central dispersion curve close to resonance. This NICE-OHMS signal is proportional to the cavity transmitted phase shift. Two theoretical plots are also shown. The blue solid line displays the full theory, i.e., including doppleron resonances to all orders, while the black dashed line is based on a theoretical prediction where the velocity dependent doppleron resonances are not taken into account. We observe very good agreement with the full theoretical model including all doppleron orders and notice the doppleron resonances tend to decrease the phase dispersion slope around resonance slightly. The two theory lines are also included in the broad frequency scan shown in Fig.~\ref{Broadscans}(a). For the dashed theory line excluding doppleron resonances, the agreement with the data is always worse than that including the doppleron resonances.

The signal obtained by demodulating the cavity transmitted field at 2FSR (Eq.~(\ref{2omega})) is shown in Fig.~\ref{Broadscans}(b) and is related to the absorption of the cavity transmitted field. A background signal corresponding to the empty cavity signal (no atoms) is subtracted. A similar background value has been subtracted from the theoretical values. The absolute value of this background corrected signal represents the degree of field absorption experienced by light transmitted through the cavity. In order to reproduce the detailed features of the absorption lineshape we must include contributions from sidebands up to the third order in the theoretical model. These higher order contributions modify the line shape. Different modulation indices of $x=0.65$ for the blue theory line and of $x=0.57$ for the green theory line are chosen to show that the signal described in Eq.~(\ref{2omega}) takes the optical power ratio of the carrier and the sidebands into account in good agreement with the data.

The absorption line shapes in Fig.~\ref{Broadscans}(b)  are shown for two different carrier input powers, $310\,\textrm{nW}$ (red dots) and $155\,\textrm{nW}$ (green dots). 
The 2FSR demodulation signal is proportional to the total optical input power. A carrier input power of $155\,\textrm{nW}$ should thus yield about half of the signal measured for a carrier input power of $310\,\textrm{nW}$. This is indeed observed experimentally, however, on the presented figure the signal for carrier input power of $155\,\textrm{nW}$ has been normalized to the signal for a carrier input power of $310\,\textrm{nW}$ in order to illustrate the line shape dynamics on the same figure.

As mentioned earlier the cavity is forced on resonance with the probe carrier by the cavity lock feedback loop such that the combined atom-cavity system is kept on resonance. However, the presence of several sideband orders contributes to the servo error signal for large atomic phase shifts and will slightly offset the cavity resonance compared to the carrier frequency \cite{NICEOHMS3}. Corrections due to these shifts are small but they are included in the theoretical model for completeness.	

The central region of Fig.~\ref{Broadscans}(b) shows a small central peak with reduced absorption (increased transmission) due to saturation. Insets show a zoom of the phase and the cavity transmitted signal in Fig.~\ref{Broadscans}(a) and ~\ref{Broadscans}(b) respectively. A reduction in transmission occurs for small detunings and forms dips on both sides of the central saturation peak at around 100 kHz. This reduction is due to atom-induced cavity response, where a non-zero atomic phase shifts the sidebands out of cavity resonance and reduces the transmitted power. As the detuning increases to larger values the phase shift induced by the atoms eventually decreases to zero again bringing the sidebands back into resonance with the cavity mode and thus increasing the amount of transmitted light. This corresponds to the two shoulders located next to the dips. This structure is less pronounced when measuring the total cavity transmitted power as a function of the detuning since the carrier, which is kept on resonance, accounts for most of the transmitted power for low modulation indices ($x<1$). On the other hand, this structure is significantly more pronounced when the cavity transmitted field is demodulated with 2FSR (Eq.~(\ref{2omega})). A detailed account of this structure can be found in Appendix~\ref{appendix}. The atom-induced cavity response seems even more pronounced in the data (see inset in Fig.~\ref{Broadscans}(b)), than predicted by theory, which might be due to an imperfect cavity locking scheme. 

The shot noise limited linewidth of the system, $\Delta\nu$, can be estimated by assuming a perfect locking scheme, that is of infinite bandwidth, and detectors with unity quantum efficiencies as \cite{MikeMartin}
\begin{eqnarray}
\Delta \nu = \frac{\pi \hbar \nu}{2 P_{\textrm{\tiny{sig}}} (\frac{d \phi}{d\nu})^2} \left(1+\frac{P_{\textrm{\tiny{sig}}}}{2P_{\textrm{\tiny{sideband}}}}\right),\label{SNL}
\end{eqnarray}
where $P_{\textrm{\tiny{sig}}}$ is the input carrier power, $P_{\textrm{\tiny{sideband}}}$ is the first order sideband power and $\frac{d \phi}{d\nu}$ is the dispersion slope at resonance. 
The minimum achievable shot noise limited linewidth of the current system is estimated from Eq.~(\ref{SNL}) to be $500\,\textrm{mHz}$. 
The minimum shot noise limited linewidth will be further reduced for higher values of $C=NC_{0}$. Hence, it can be optimized by increasing the atom number or the cavity finesse.

A system with improved finesse ($F=1000$) and intra cavity atom number ($N_{\textrm{\tiny{cav}}}$) increased by a factor 2 is estimated to have a shot noise limited linewidth of $\Delta\nu<10\,\textrm{mHz}$ \cite{DT}, which is comparable to state-of-the-art laser stabilization \cite{Kessler,Bishof}. 
Furthermore, the theoretical model shows that the gain, in terms of dispersion slope $\frac{d \phi}{d\nu}$, by further cooling of the atomic sample is minimal \cite{PGW}. Experimentally, the reduction of the sample temperature by one or two orders of magnitude in temperature is usually accompanied by a reduction in atom number by similar orders of magnitude. In this regime with lower temperature the optimum input power is reduced to technically challenging values ($<\textrm{nW}$). 
This potentially opens prospects for implementations of thermal atoms in a simple and compact transportable optical atomic clock with high phase stability.

\begin{figure}[h!t]
\includegraphics[width=1.00\columnwidth]{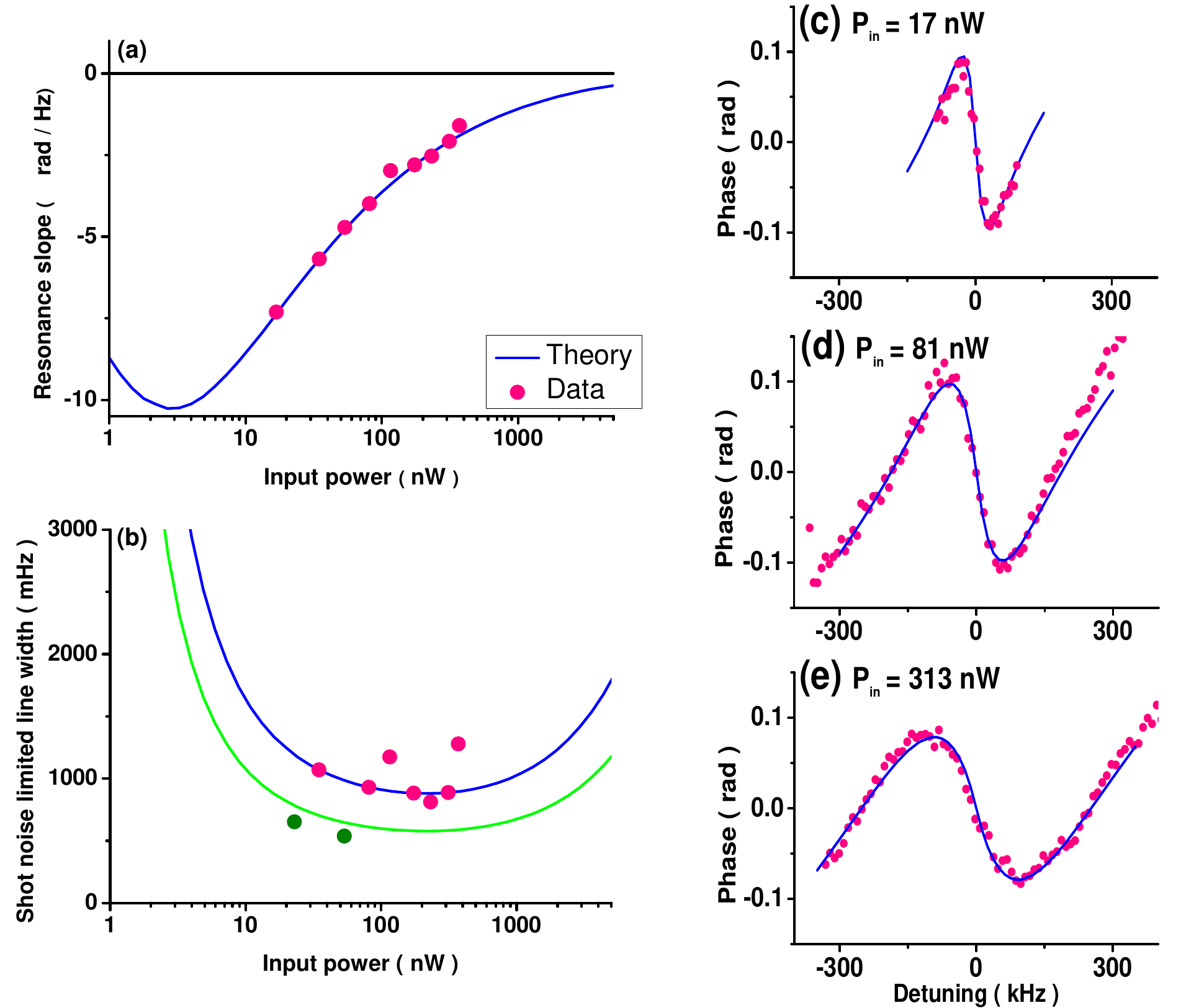}
\caption{The input power dependence of the slope of the phase dispersion around resonance and the corresponding shot noise limited linewidth for a total number of atoms of $N=4.0\cdot 10^{8}$ and MOT temperature of $T=5.0\,\textrm{mK}$. The solid lines are theoretical predictions and the dots are experimental data. \textbf{(a)} The slope of the phase dispersion around resonance measured for different input powers. The data and theory are in excellent accordance and the non-linear dependence is evident. Notice that the power-axis is logarithmic. \textbf{(b)} The shot noise limited linewidth from Eq.~(\ref{SNL}) for different input powers corresponding to the measured parameters in (a). The solid lines are theoretical predictions and the dots are experimental values. The blue theory line corresponds to a ratio of carrier and sideband input powers of $\frac{P_{\textrm{\tiny{sig}}}}{2P_{\textrm{\tiny{sideband}}}}=1.65$ and the green theory line corresponds to $\frac{P_{\textrm{\tiny{sig}}}}{2P_{\textrm{\tiny{sideband}}}}=0.75$.
\textbf{(c)-(e)} Frequency scans of the phase dispersion for different input powers. All experimental values in (a) and (b) are evaluated by fitting theoretical curves to frequency scans of the phase dispersion around resonance. The non-linear dynamics of the overall dispersion shape is in very good accordance with the theory.
\label{slopeplots}}
\end{figure}

The phase dispersion slope depends non-linearly on the optical intra-cavity power, which makes the optimal choice of parameters for laser stabilization non-trivial. The non-linear input power dependence of the phase dispersion slope is measured and shown in Fig.~\ref{slopeplots}(a). The slope values are evaluated by performing theoretical fits on phase dispersion scans, shown in Fig.~\ref{slopeplots}(c)-(e), for different input powers. The measurements and theory are in good agreement and the non-linear input power dependency is evident. The theoretical curve predicts a maximum absolute value of the slope for lower input powers than measured. However, the shot noise limited linewidth does not depend solely on the phase dispersion slope. It is also governed by the input power in accordance with Eq.~(\ref{SNL}). The shot noise limited linewidth for each measured parameter is calculated and shown in Fig.~\ref{slopeplots}(b), where data sets in Fig.~\ref{slopeplots}(a) and (b) are identical. Figure~\ref{slopeplots} (b) shows, that a system with optimum input power for a fixed carrier-sideband ratio has been realized and measured. Note that the horizontal axis is logarithmic, allowing a change in input power of one order of magnitude without degrading the shot noise limited linewidth significantly. This indicates that probing with technically challenging low input powers down to $<100\,\textrm{nW}$ is not necessary, and that the system is robust to changes in the input power. The ratio of carrier and sideband powers is governed by the modulation index $x$ of the EOM, and optimizes the linewidth in Eq.~(\ref{SNL}) for 
a carrier-sideband ratio of $\frac{P_{\textrm{\tiny{sig}}}}{2P_{\textrm{\tiny{sideband}}}}=1$.
The measurements presented in this work are performed in a cyclic manner, and continuous laser stabilization on the presented system is not currently feasible. However, the implementation of the techniques presented in this work into beam-line experiments \cite{Chen}, where a beam of cold atoms with high loading rates \cite{Wilkovsky} are interrogated in a cavity, seems realizable.
This opens the possibilities for a continuous laser stabilization on an atom-cavity system with a narrow optical transition.

\section{conclusion}

We have experimentally investigated the velocity dependent spectroscopic features and non-linear dynamics of a cavity-atom system in the bad cavity limit consisting of laser-cooled strontium-88 atoms coupled to a low finesse optical cavity.

The NICE-OHMS technique has been applied to perform FM spectroscopy and a complete connection is established between the measured quantities and the complex transmission coefficient of the input field. The cavity transmitted phase shift and the absorption of the cavity transmitted field are measured and both quantities show significant modifications due to velocity dependent processes and atom-induced cavity response in accordance with the theoretical model.

The ideal shot noise limited linewidth of a laser stabilized to our system depends on the phase dispersion slope around resonance. The non-linear input power dependence of the phase dispersion slope is measured and the optimal input power for minimum shot noise limited linewidth is determined. Parameters corresponding to a shot noise limited linewidth down to $500\,\textrm{mHz}$ are measured.

The understanding of relevant velocity dependent effects presented here has direct relevance for atomic clocks and superradiant laser sources \cite{Meiser1} involving ultra narrow transitions.
Specifically, the understanding of the dynamics of an cavity-atom system with single-stage MOT temperatures obtained in this work will prove valuable for future transportable stable lasers and atomic clocks employing thermal atoms for out-of-lab operation under more noisy environments.\newline

\begin{acknowledgments}
The authors would like to acknowledge support from the Danish research council
and ESA contract number No. 4000108303/13/NL/PA-NPI272-2012. JY and MH also wishes to thank the DARPA QuASAR program, the NIST and the NSF for financial support.
\end{acknowledgments}

\appendix
\section{Line shape of absorption}\label{appendix}

The absorption dips shown in the inset of Fig.~\ref{Broadscans}(b) are predicted from our theoretical model \cite{PGW,DT}, where the empty cavity frequency is kept on resonance with the probe field. The standing wave condition inside the cavity is however unfulfilled when a strong atomic dispersion is introduced, and the transmission will be reduced for all detunings with non-zero atomic phase shifts. Hence the large enhancement of the absorption for large detunings ($\sim\textrm{MHz}$) is also due to the broad thermal dispersion.

Experimentally, the cavity servo system ensures that the combined resonance of the cavity and the atoms is kept on resonance with the probe field. The standing wave condition is hence maintained and the atomic phase shift is canceled out for the carrier as shown in Eq.~(\ref{resonancecondition}). A measurement of the transmission power of the system will be dominated by the complex transmission coefficient of the carrier, $\chi_{0}$ (see Eq.~(\ref{Chi0})), and the change in the transmission power due to the atomic phase shift is not evident. The signal demodulated with 2FSR (Eq.~(\ref{2omega})), on the other hand, contains information about the atomic absorption ($\chi_{0}$) and the attenuation due to the atomic phase shift ($\Re(\chi_{2})$). Hence, the detailed absorption structure is evident in this signal.

\begin{figure}[ht]
\includegraphics[width=1.0\columnwidth]{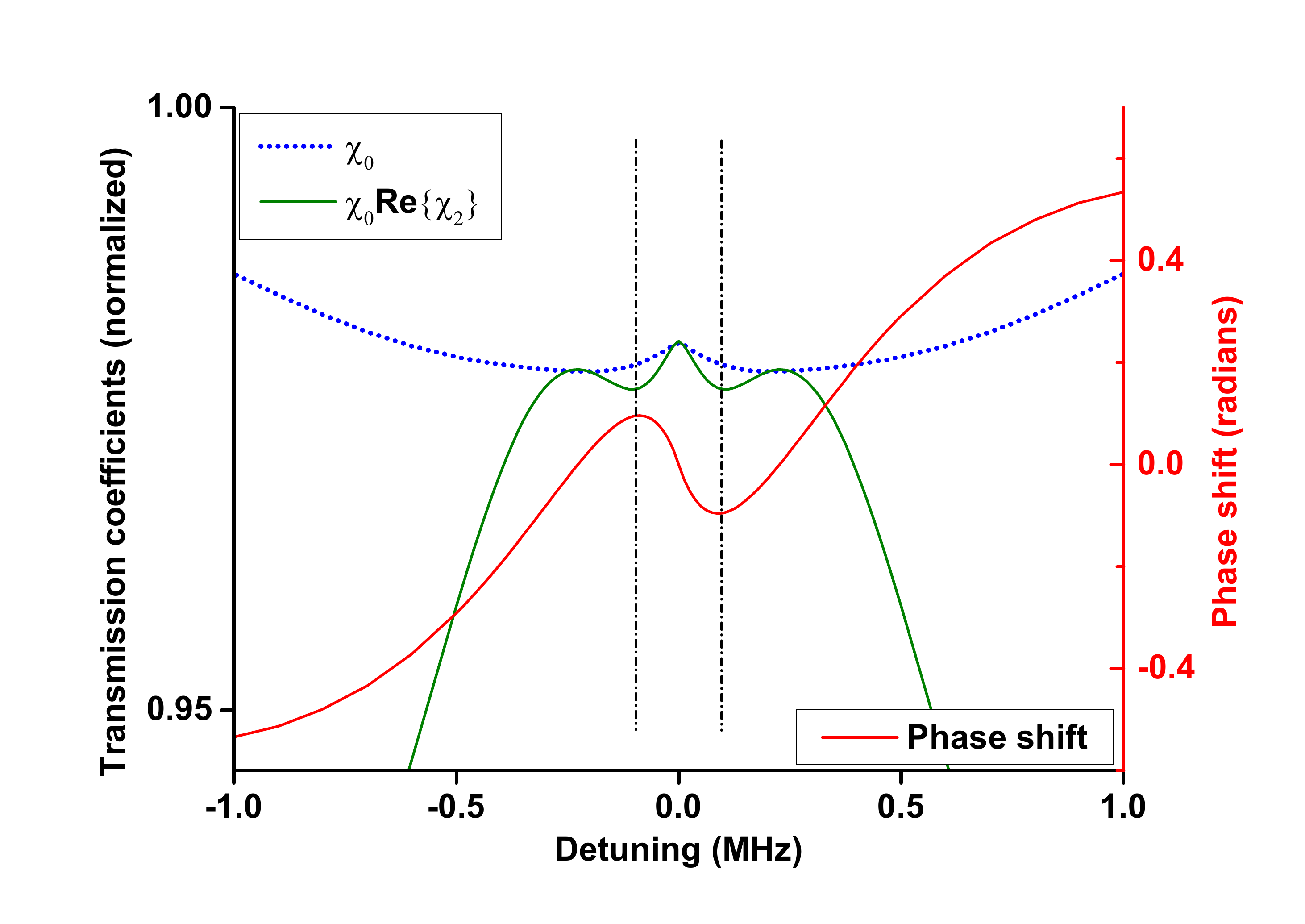}
\caption{Theoretical values of the transmission coefficient of the carrier (dashed blue), $\chi_{0}$, the combined line shape of the atomic absorption and the attenuation due to atomic dispersion (green), $\chi_{0}\Re(\chi_{2})$, and the total cavity transmitted phase shift (red). The black vertical dashed lines highlights that transmission minima occur at dispersion maxima.
\label{appendixfigure}}
\end{figure}

Theoretical values of some relevant transmission coefficients are shown in Fig.~\ref{appendixfigure}. $\chi_{0}$ (blue line) describes only the atomic absorption of the carrier (see Eq.~(\ref{Chi0})), $\chi_{0}\Re(\chi_{2})$ describes the combined line shape from the atomic absorption ($\chi_{0}$) and the attenuation due to the atomic phase shift ($\Re(\chi_{2})$). $\chi_{0}\Re(\chi_{2})$ corresponds to the first term of Eq.~(\ref{2omega}) measured by the NICE-OHMS technique with demodulation of 2FSR. The second term in Eq.~(\ref{2omega}) and the additional contributions from the higher order sidebands included in the theoretical description modify the line shape without erasing the structures.
The corresponding total dispersion of the system is also shown in Fig.~\ref{appendixfigure}, and it is evident that the transmission minima occurs at the dispersion maxima (highlighted with dashed vertical lines).

\end{document}